\documentclass[times,twocolumn,nopreprintline,final,fleqn]{elsarticle}

\usepackage{framed,multirow}
\RequirePackage{graphicx}

\usepackage{amsfonts,amssymb,amsmath}
\usepackage{soul}

\soulregister{\cite}7 
\soulregister{\citep}7 
\soulregister{\citet}7 
\soulregister{\ref}7 
\soulregister{\pageref}7 

\usepackage{url}
\definecolor{miacolor}{RGB}{0,128,172}  
\usepackage{multirow}
\usepackage{booktabs}
\usepackage{array} 
\usepackage{threeparttable}

\usepackage{hyperref}
\hypersetup{hypertex=true,
    colorlinks=true,
    linkcolor=ccr,
    anchorcolor=ccr,
    citecolor=ccr}

\usepackage[utf8x]{inputenc} 

\definecolor{newcolor}{rgb}{.8,.349,.1}

\begin{document}


\begin{frontmatter}

\title{Neural Implicit Surface Reconstruction of
 Freehand 3D Ultrasound Volume 
with Geometric Constraints}%

\author[1,2,3]{Hongbo Chen}

\address[1]{{School of Information Science and Technology, ShanghaiTech University},
    {Shanghai},
    {201210}, 
    {China}}

  \address[2]{{Shanghai
  Advanced Research Institute, Chinese Academy of Sciences},
  {Shanghai},
  {200050}, 
  {China}}

\address[3]{{University of Chinese Academy of Sciences},
  {Beijing},
  {100049}, 
  {China}}

\author[4]{Logiraj Kumaralingam}

\author[1]{Shuhang Zhang}

\author[1]{Sheng Song}

\author[1]{Fayi Zhang}

\author[1]{Haibin Zhang}

\author[4,5]{Thanh-Tu Pham}

\author[4]{Kumaradevan Punithakumar}

\author[4,5]{Edmond H. M. Lou}

\author[1]{Yuyao, Zhang}

\author[4,5]{Lawrence H. Le\corref{cor1}}
\ead{lawrence.le@ualberta.ca}

\author[1,6]{Rui Zheng\corref{cor1}}
\ead{zhengrui@shanghaitech.edu.cn}

\address[4]{{Department of Radiology and Diagnostic Imaging, University of Alberta},
    {Edmonton},
    {T6G 2V2}, 
    {Alberta},
    {Canada}}

\address[5]{{Department of Biomedical Engineering, University of Alberta},
    {Edmonton},
    {T6G 2V2}, 
    {Alberta},
    {Canada}}

\address[6]{{Shanghai Engineering Research Center of Intelligent Vision and Imaging, ShanghaiTech University},
    {Shanghai},
    {201210}, 
    {China}}

\cortext[cor1]{Corresponding author}




\begin{abstract}
Three-dimensional (3D) freehand ultrasound (US) is a widely used imaging modality 
that allows non-invasive imaging of medical anatomy without radiation exposure. 
Surface reconstruction of US volume is vital to 
acquire the accurate anatomical structures needed for 
modeling, registration, and visualization.
However, traditional methods cannot produce a 
high-quality surface due
to image noise.
Despite improvements in smoothness, continuity, and resolution 
from deep learning approaches, 
research on surface reconstruction in 
freehand 3D US is still limited. 
This study introduces FUNSR,
a self-supervised neural implicit surface 
reconstruction method to learn signed distance functions (SDFs) 
from US volumes. 
In particular, FUNSR iteratively learns the SDFs by moving 
the 3D queries sampled 
around volumetric point clouds to approximate the surface,
guided by two novel geometric constraints:
sign consistency constraint and on-surface constraint with adversarial learning.
Our approach has been thoroughly evaluated across four datasets to 
demonstrate its adaptability
to various anatomical structures,
including a hip phantom dataset, two vascular datasets and one publicly available prostate dataset.
We also show that smooth 
and continuous representations greatly 
enhance the visual appearance of US data. 
Furthermore, we highlight the potential of our method 
to improve segmentation performance, and its robustness to noise distribution and motion perturbation.

\end{abstract}


\end{frontmatter}


\section{Introduction}
Ultrasound (US) is a widely used imaging modality for clinical diagnosis,
monitoring and analysis due to its low cost,
radiation-free and real-time performance. 
The 3D US with rich anatomical structure has been greatly 
developed and applied in many areas,
such as 3D carotid artery diagnoses, automatic segmentation and organ visualization
~\citep{jiangDualstreamCenterlineguidedNetwork2023,liuDeepLearningMedical2019a,nambureteNormativeSpatiotemporalFetal2023}.
The data acquisition of 3D US imaging 
can be divided into 2D array, mechanical control, 
freehand scan without tracking sensor and 
freehand scan with tracking sensors~\citep{pragerThreedimensionalUltrasoundImaging2010,prevost3DFreehandUltrasound2018a, luoDeepMotionNetwork2022a,guoUltrasoundVolumeReconstruction2022}.
Compared to the other three approaches,
freehand 3D US with tracking sensors is a rapidly advancing 
technology to obtain the high-quality 3D volumes
without limitation of field-of-view~\citep{rohlingComparisonFreehandThreedimensional1999,mozaffariFreehand3DUltrasound2017}.


Freehand 3D US imaging technology commonly 
compounds the collected 2D B-mode images 
from transducer and the corresponding 3D poses from tracking device
into a volume
to reconstruct 3D internal structures of human body
~\citep{mozaffariFreehand3DUltrasound2017}. 
After reconstruction, volume rendering technique is usually
adopted to transmit the 3D information in a translucent way or 
implicit neural representation~\citep{li3DUltrasoundSpine2021,9958448,wysockiUltraNeRFNeuralRadiance2023a}.
However, such voxel-based visualization makes it hard to recover accurate geometries
and continuous surface~\citep{batlleLightNeuSNeuralSurface2023}.
Since each voxel element in the volume is considered for the rendered view,
this technique also requires expensive computation costs for training~\citep{huangReviewRealTime3D2017b,9958448}.

Surface reconstruction is an 
alternate method to 
facilitate the geometric evaluation, morphological assessment, anatomical structure visualization, 
and surgery guidance
~\citep{zhouRealTimeDenseReconstruction2020,yeungLearningMap2D2021a,velikovaImplicitNeuralRepresentations2024,mengDeepMeshMeshBasedCardiac2024}.
Traditional surface reconstruction methods for freehand 3D US,
like Contour filtering and Marching Cubes,
usually
convert the voxel representation after volume reconstruction
to triangles or polygons mesh according to the segmented boundary
~\citep{mohamedSurvey3DUltrasound2019}.
The geometric connection between each voxel is established following
successive slices or voxel intensity, 
also known as ISO-Surface~\citep{treeceRegularisedMarchingTetrahedra1999}. 
However, traditional methods have connectivity problems because of 
the reconstructed empty holes 
or noise in US volume, 
this will lead to a rough surface, which requires further post-processing, 
such as smoothing and interpolation for better reconstruction results~\citep{zhangSurfaceExtractionThreedimensional2004}.
The quality of generated surface is also limited by 
the original reconstructed volume and 
voxel resolution~\citep{nguyenValidation3DSurface2015a}. 

In recent years, there has been a growing interest in using deep learning (DL) methods 
for surface reconstruction from medical imaging data, 
as an alternative to traditional methods, 
to bring increased accuracy and speed, as well as the ability to handle more 
complex and heterogeneous data~\citep{farshianDeepLearningBased3DSurface2023}.
Among various approaches, implicit neural representation (INR) has received substantial attention 
in recent few years and successfully applied in medical imaging and graphics
~\citep{molaeiImplicitNeuralRepresentation2023a,reedNeuralVolumetricReconstruction2023}.
INR typically parameterizes a 3D surface structure as neural 
implicit functions 
through a deep neural network, such as encoding the signed 
distance functions (SDFs) as Multi-Layer Perceptrons (MLP)~\citep{parkDeepSDFLearningContinuous2019}.
Within this context, INR provides resolution-agnostic 
surface rendering and improved memory efficiency. 
After training, the
3D scene can be effectively represented by a straightforward
MLP, with the flexibility to re-generate mesh at varying 
resolutions during the reference stage.

However, the application of INR to 
freehand 3D US for
continuous surface reconstruction 
is still missing.
There are two main challenges in 3D freehand US to achieve 
high-quality surface reconstruction
using deep learning.
1) Only the boundary of target structure in 2D segmentation mask
cannot accurately reflect actual smooth surface 
because of the low signal-to-noise ratio (SNR) and poor pixel connectivity in US images.
Additionally, it's difficult to extract the exact border and normal of complex surface. 
2) Different anatomical structures have different geometric priors, and,
it's difficult to acquire enough ground truth volumes, templates or surface normal to 
supervise the training stage for 3D surface reconstruction.

In this study, in order to address these issues,
we propose, FUNSR, an efficient end-to-end online learning model
to predict the SDFs for implicit surface representation 
 from freehand 3D US point clouds data. 
Instead of preprocessing and extracting segmentation boundary as surface point cloud, 
the whole segmentation masks
and their locations are directly transformed into a 3D volumetric 
point cloud as the input. 
Our target is to use the neural network to overfit underlying 
geometric surface
in point clouds as the continuous SDFs representation to 
avoid disturbance of noise.
Specifically, a set of 3D query points is sampled around each point in point clouds,
and fed into the network to learn SDFs between query points and point clouds
using a self-supervised strategy~\citep{NeuralPull}.

In summary, there are four primary contributions: 
\begin{itemize}

  \item[$\bullet$] To the best of our knowledge, this is the first attempt 
  to introduce 
  an online self-supervised learning strategy of
  neural implicit surface reconstruction for freehand 3D US volumetric point clouds.
  The proposed method can learn 3D continuous and high-resolution structures 
  from individual input subject 
  without the need for additional ground truth training data or post-processing.
  
  \item[$\bullet$] We introduce two geometric constraints  
  aimed at enhancing surface reconstruction accuracy.
  
  1) Sign consistency constraint which is designed to correct the predicted SDFs 
  within volumetric point clouds. 
  This constraint ensures the network maintains an accurate sign 
  in the input mask. 

  2) On-surface constraint  
  in accordance with adversarial learning strategy which
  is adopted to dynamically optimize 
  the predicted SDFs when approaching to 
  zero-level-set surface.
  
  \item[$\bullet$] 
  We conduct extensive experiments to assess the structural fidelity and computational efficiency of 
  our approach using four distinct datasets acquired by different US imaging systems.
  These datasets encompass four types of anatomical structures:
  hip joint, common carotid artery, carotid artery bifurcation and prostate (public dataset).
  The results indicate that our approach
  is competitive with current methods.

  \item[$\bullet$] 
  We demonstrate the robustness of our model against various categories of input noise, 
  including network 
  segmentation noise, randomly distributed outliers and 
  unexpected motion perturbation during freehand operation. 
  We also thoroughly conduct ablation studies to investigate the effects of  
  introduced geometric constraints and two input point cloud modalities on
  the surface reconstruction. 
  We show that directly using volumetric point cloud is more straightforward and
  produces higher reconstruction quality compared to surface point cloud.
  
\end{itemize}

\section{Related works}

\subsection{Traditional surface reconstruction in 3D ultrasound}
Traditional surface reconstruction methods in 3D US can be simply 
classified into direct extraction and indirect segmentation.
\cite{zhangDirectSurfaceExtraction2002,zhangSurfaceExtractionThreedimensional2004} propose 
to directly extract the surface from freehand 3D US B-mode scan using 
ISO-Surface after approximating a radial basis function
across the pixels in 2D transverse images. 
\cite{nguyenValidation3DSurface2015a} first segments the contours of region 
of interest (ROI) from
transverse B-mode images in
the reconstructed US volume.
The Bézier-spline function is then used
to smooth and interpolate the contours to form a high-quality triangle mesh. 
\cite{kerrAccurate3DReconstruction2017} adopts
a column-wise threshold method to segment the bony surface
of each image as a point cloud from ultrasonic synthetic aperture images.
Surface mesh is directly produced by the point clouds
using a wrapping algorithm.

\subsection{DL-based surface reconstruction in medical imaging}
Recent advances in DL-based 3D surface reconstruction 
methods have shown the 
promising results in magnetic resonance image(MRI), 
computerized tomography(CT) and 3D echocardiography.
\cite{wangShapeReconstructionAbdominal2021} extends the graph 
convolutional network (GCN) 
to learn mesh deformation parameters of the 
target CT organ from surrounding
detectable-organ features. 
A target organ template is provided to receive the predicted deformation parameters.
After combining convolutional neural network (CNN) and GCN,
the network can reconstruct 3D liver surface from a single 
2D radiograph~\citep{nakaoImagetoGraphConvolutionalNetwork2021}.
Strategies for joint segmentation and surface reconstruction have been designed 
to facilitate end-to-end training,
converting
the input MRI or CT volume directly to output meshes
~\citep{gopinathSegReconLearningJoint2021,wickramasingheVoxel2Mesh3DMesh2020}.
Cortical surface reconstructions from brain MRI scans mainly focus on 
the development of deformable template-based geometric networks~\citep{santacruzCorticalFlowBoostingCortical2022,maCortexODELearningCortical2023,bongratzNeuralDeformationFields2024}.
For echocardiography, \cite{LAUMER2023102653} proposes a weakly supervised auto-encoder model  to generate a 
4D shape mesh  
from sequences of 2D mesh videos.

\subsection{Implicit neural representation}

\textit{Neural volume reconstruction}. 
Concept of INR
has been widely applied in different medical imaging modalities for volume reconstruction, 
including MRI slice-to-volume reconstruction~\citep{xuNeSVoRImplicitNeural2023}, 
CT limited-view reconstruction~\citep{songPINERPriorInformedImplicit2023},
3D US freehand (sensorless) volume reconstruction~\citep{wysockiUltraNeRFNeuralRadiance2023a,yeungSensorlessVolumetricReconstruction2024,gaitsUltrasoundVolumeReconstruction2024,eidRapidVolRapidReconstruction2024}.
In particular, \cite{velikovaImplicitNeuralRepresentations2024} 
attempts to integrate the neural volume reconstruction (NVR) and a post-processing module
to reconstruct the abdominal aorta surface
from robotic 3D US imaging. 
This method (NVR-Poisson) first extracts a dense point cloud from the neural reconstructed
volume, 
then downsamples the point cloud and fits a convex hull to obtain surface point cloud and corresponding point normal.
The mesh is finally obtained through Poisson surface reconstruction.

\textit{Neural surface reconstruction}.
Currently, the development of end-to-end learning of an SDF 
has emerged as a primary choice for neural implicit 
surface reconstruction~\citep{zhaEndoSurfNeuralSurface2023,batlleLightNeuSNeuralSurface2023}.
\cite{parkDeepSDFLearningContinuous2019} 
proposes DeepSDF to learn an SDF in continuous space with a latent code 
by given ground truth signed distance values.
\cite{NeuralPull} further demonstrates the 
feasibility of SDFs to learn 
3D shape from surface point clouds using
a neural pull operation 
without extra ground truth signed distance values, point
cloud normal or occupancy value. 
\cite{cruzDeepCSR3DDeep2021} introduces the INR to MRI cortical surface reconstruction
for the points in a continuous brain template coordinate system.
Inspired by DeepSDF, \cite{sanderReconstructionCompletionHighresolution2023}
trains an auto-decoder architecture to complete the high-resolution 3D left ventricle shape from 
sparse views of cardiac MRI (CMRI).
Furthermore, \cite{wiesnerGenerativeModelingLiving2024} encodes the spatial domain with an additional temporal domain 
into network to learn the shape of living cells from microscopy.
Alongside SDFs, 
a recent approach has explored representing high-resolution shape through learning implicit
occupancy function~\citep{meschederOccupancyNetworksLearning2019} using the architecture of DeepSDF from 
sparse volumetric masks in CT and MRI scans~\citep{amiranashviliLearningContinuousShape2024b}. 
In contrast to this occupancy-focused work, our study aims to 
reconstruct geometric surface and simultaneously establish a signed distance field, 
enabling potential applications in surgical navigation and multi-view reconstruction~\citep{driessLearningModelsFunctionals2021}.

\begin{figure*}[t]
	\centering
	\includegraphics[width=15cm]{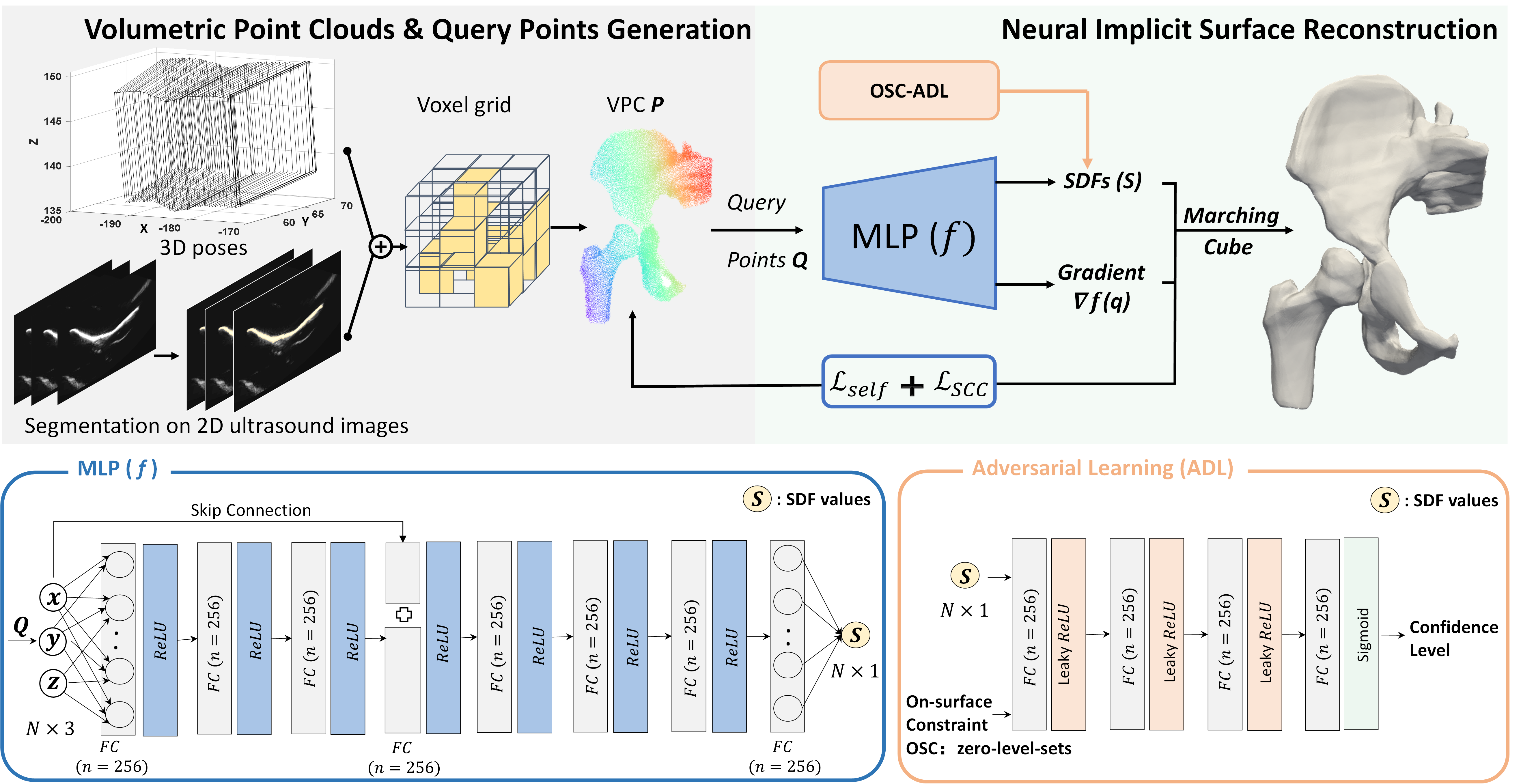}
	\caption{Overview of the proposed FUNSR, a neural implicit surface reconstruction method from 
  freehand US volumetric point clouds (VPC).
  The VPC $\boldsymbol{P}$ is first extracted from freehand 3D US volume after segmentation.
  A series of query points $\boldsymbol{Q}$ are then sampled around each point in $\boldsymbol{P}$ and fed into the neural network
  for learning of gradient ($\nabla f\left(q\right)$) via backpropagation and SDFs ($\boldsymbol{s}$).
  The final mesh file is generated using Marching Cube algorithm using the predicted SDFs.
  Blue block at the bottom left depicts the MLP neural network with 
  fully connected layers (FC), 
  and orange block at the bottom right represents the 
  on-surface constraint with adversarial learning (OSC-ADL).  }
	\label{Figure1}
\end{figure*}

\section{Materials and Methods}

\subsection{Overview} 
\label{Section_Overview}

Our proposed method consists of two primary components:
volumetric point cloud $\&$ query points generation, and neural implicit surface reconstruction.
In addition, two geometric constraints, including sign consistency constraint (SCC) and 
on-surface constraint with adversarial learning (OSC-ADL),
are designed to ensure superior reconstruction quality.

As shown in Fig.~\ref{Figure1}, given a sequence of 2D transverse images with associated 3D freehand poses, 
the coordinates of 2D segmented masks
are transformed into a 3D point cloud $\boldsymbol{P}$ in 
 tracking space. 
The query points $\boldsymbol{Q}$ are then sampled
around each point in $\boldsymbol{P}$ as input of MLP network. 
The SDFs between $\boldsymbol{Q}$ and underlying surface 
described by $\boldsymbol{P}$ are then progressively learned
via neural network, guided by geometric constraints in a self-supervised manner.
The triangle mesh is ultimately produced 
by Marching Cube algorithm using the predicted SDFs~\citep{lorensenMarchingCubesHigh1987}.

\begin{figure*}[h]
	\begin{center}
	\includegraphics[width=12cm]{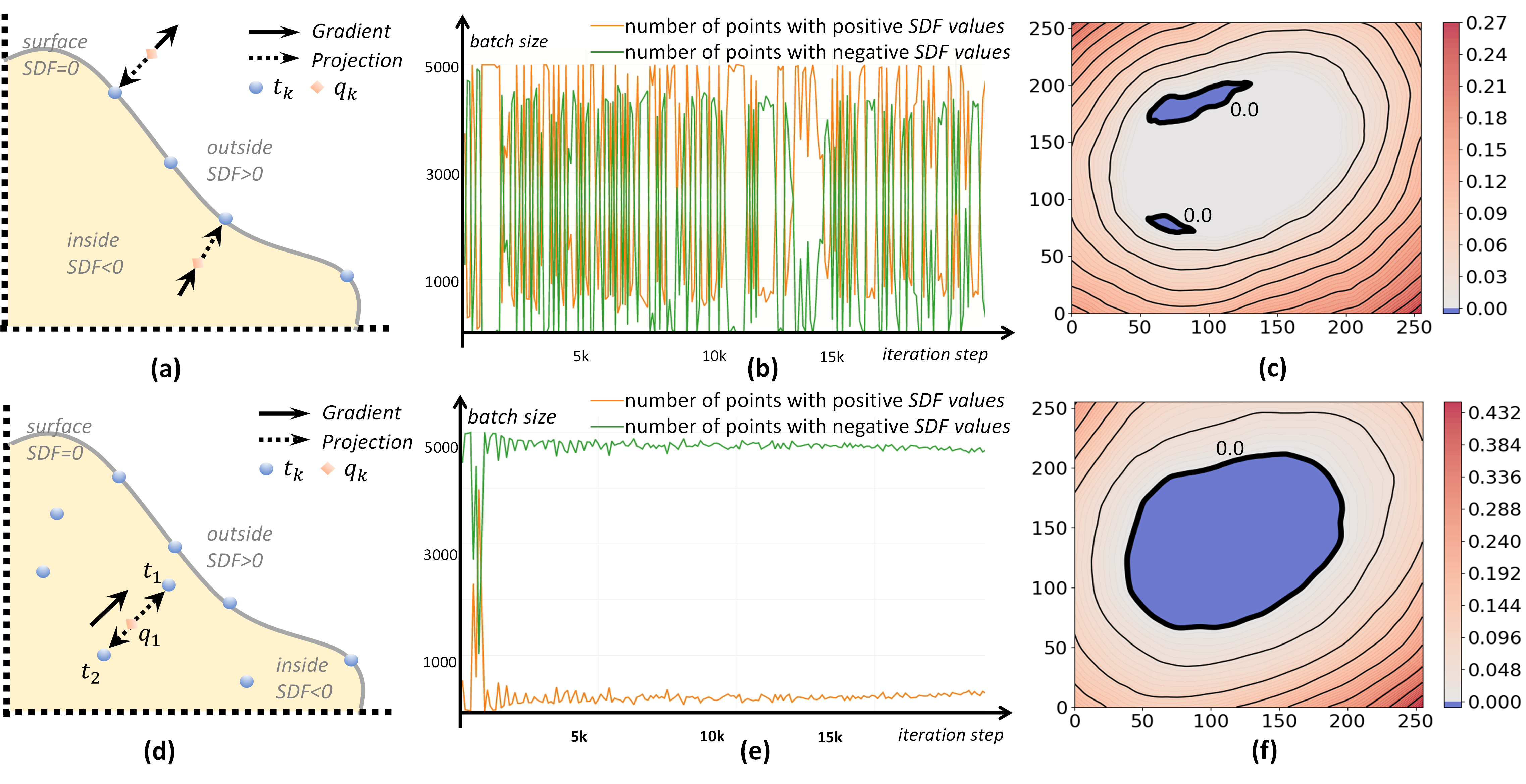}
	\caption{The 2D demonstration of sign consistency constraint from carotid sample. 
    (a) and (d) Projection around the surface and inside surface. 
    (b) and (e) Plot of the number of points with pos/neg SDF value learned from $\mathcal{L}_{self}$  loss only and with SCC. 
    (c) and (f) Level-sets of SDF learned from $\mathcal{L}_{self}$ loss only and with SCC. 
    The blue region and red region indicate the negative area (inside) and positive area (outside), respectively.
  }
	\label{Figure2}
  \end{center}
\end{figure*}

\subsection{Volumetric point cloud and query point generation} 
\label{Section_VPC_generation}

For freehand 3D US system~\citep{chenImprovement3DUltrasound2021}, each 2D pixel $\boldsymbol{v_s}$ in the segmented mask space 
is first transformed into tracking space to 
acquire the 3D position, $\boldsymbol{v_t} = M_tM_c\boldsymbol{v_s}$.
Where, $\boldsymbol{v_s} = [x_s,y_s,0,1]^T$ and $\boldsymbol{v_t} = [x_t,y_t,z_t,1]^T$ are in homogeneous coordinate representation.
$M_c$ and $M_t$ are the $4\times4$ calibration and 
tracking matrices from imaging system.

These 3D points $\boldsymbol{v_t}$ are then spatially discretized 
into voxel grids and downsampled to 
acquire the volumetric point cloud 
$\boldsymbol{P}=\left\{\boldsymbol{p_i}, i = 1,...,N \right\}$ 
for uniform 
density distribution and efficient memory usage
~\citep{zhaoLearningAnchoredUnsigned2021,kerrAccurate3DReconstruction2017}.
For each $\boldsymbol{p_i}$ in $\boldsymbol{P}$, a set of 3D query points $\boldsymbol{Q}=\left\{\boldsymbol{q_{ij}}, j = 1,...,l \right\}$
 are randomly sampled around it
following Gaussian distribution $(0,\delta)$~\citep{NeuralPull}.
Here, $\delta$ is selected as the Euclidean distance between $\boldsymbol{p_i}$ and its $k$-th nearest neighbor,
${N}$ and ${l}$ are the total number of  
3D point cloud and query points, respectively.

\subsection{Neural implicit surface reconstruction}
\label{Section_NISR}

\subsubsection{Self-supervised learning of SDFs from point clouds}

The SDFs are an implicit continuous representation to describe 
 3D shape by dividing 
the surface as exterior (positive), on-surface (zero) and interior 
(negative) ~\citep{zhaEndoSurfNeuralSurface2023, parkDeepSDFLearningContinuous2019}.
In this paper, we indicate the SDFs through learning 
signed distances between 
 surface represented by point cloud 
$\boldsymbol{P}$ 
and query points $\boldsymbol{Q}$:

\begin{equation}
  SDF\left(\boldsymbol{q}_{i j}, \boldsymbol{p}_{\boldsymbol{i}}\right)=
\left\{\boldsymbol{s} \in \mathbb{R} \:|\: 
\boldsymbol{q}_{i j} \in \mathbb{R}^3, 
\boldsymbol{p}_{i} \in \mathbb{R}^3 
\right\}
\end{equation}
where $\boldsymbol{p_i}= [x_i,y_i,z_i]$ is the 3D point in volumetric 
point cloud,
$\boldsymbol{q_{ij}}=[x_{ij},y_{ij},z_{ij}]$ is sampled 
query point around $\boldsymbol{p_i}$, 
and $\boldsymbol{s}$ is the predicted
signed distance between $\boldsymbol{q_{ij}} \in \boldsymbol{Q}$ and 
surface represented by $\boldsymbol{p_i} \in \boldsymbol{P}$.
The underlying surface boundary is depicted by 
a zero-level-set of $SDF(\boldsymbol{*})=0$,
which can be easily used to extract mesh by Marching Cube.

To simplify the notation, each query point is re-denoted as 
$\left\{\boldsymbol{q_k}, k = 1,...,l \right\}$.
Then, for each $\boldsymbol{q_k}$, 
we find its nearest neighbor point $\boldsymbol{t_k} \in \boldsymbol{P}$ 
under the Euclidean distance for calculation of self-supervised loss.
As described in Fig.~\ref{Figure1}, an MLP neural network $\boldsymbol{f}$ is trained to
learn the SDF value and gradient of a given query 
point $\boldsymbol{q_k}$ in $\boldsymbol{Q}$.
The query point is then projected along or against the learned gradient
to its nearest point $\boldsymbol{t_k}$ using the predicted SDF value.
This projection process is illustrated in Fig.~\ref{Figure2} (a) and formulated as the below equation~\citep{chou2022gensdf,NeuralPull},
\begin{equation}\label{EqProjection}
	{q_k}^{\prime}=q_k - \boldsymbol{s} 
  \times 
  \frac{\nabla f\left(q_k\right)}{\left\|\nabla f\left(q_k\right)\right\|_2} 
\end{equation}
where $\boldsymbol{s}$, also denoted as ${f}(q_k)$, is the predicted SDFs of $q_k$ from the MLP network, 
$\nabla f\left(q_k\right) /\left\|\nabla f\left(q_k\right)\right\|$ is 
the normalized gradient of 
network, indicating the direction in 3D space 
where signed distance increases most rapidly. 
Here, $\nabla f\left(q_k\right)$ is derived by backpropagation during the MLP training.

Following Neural-pull~\citep{NeuralPull}, we minimize 
the self-supervised L2-distance loss to optimize the projected 
${q_k}^{\prime}$ 
approaching its nearest neighbor $\boldsymbol{t_k}\in\boldsymbol{P}$:

\begin{equation}\label{Eq_Lossself}
	\mathcal{L}_{self}=\frac{1}{K} \sum_{k \in[1, K]}\left\|{q_k}^{\prime}-t_k\right\|_2^2
	\end{equation}
where $K$ is the batch size during the training.

\subsubsection{Sign consistency constraint for volumetric point clouds}

In order to learn a better underlying surface, our input is the whole mask instead 
of boundary. 
However, as proved in~\citep{chou2022gensdf}, only using Eq.~(\ref{Eq_Lossself}) 
to supervise training has no explicit penalties for the 
predicted wrong sign when query point is close to the point cloud.
This limitation will cause the network to fail to converge to a stable status,
especially inside the mask.

A simple 2D case is illustrated in Fig.~\ref{Figure2} (d),
$t_1$ and $t_2$ are two points inside the surface, $q_1$ is the corresponding nearest query point around
$t_1$ and $t_2$, respectively. 
Following Eq.~\ref{EqProjection}, the network will predict a negative sign distance to 
project $q_1$ to $t_1$ along the direction of gradient and
minimize loss Eq.~(\ref{Eq_Lossself}).
However, in next training batch, $t_2$, as the opposite direction 
of $t_1$, will make the network try to 
change the sign to minimize Eq.~(\ref{Eq_Lossself}) against the gradient.
Obviously, this will lead the predicted sign inside mask to be in a chaotic state.
As shown the Fig.~\ref{Figure2} (b) and (c), the number of positive signs and negative signs varies wildly at each iteration and
the zeros-level-set boundary disappears.

To address this problem, we introduce a direction loss as $\mathcal{L}_{SCC}$ 
to balance the distance loss by following formula,
\begin{equation}
  \mathcal{L}_{SCC}=\frac{1}{K} \sum_{k \in[1, K]} 1-\cos 
\left(\nabla f({q_k}),  \frac{({q_{k}^{\prime}} - {t_k})}{\left\|{q_k}^{\prime}-t_k\right\|_2}\right)
\end{equation}\label{Eq_lossSCC}
where $1-cos(*,*)$ is the cosine distance to penalize 
direction deviation between two non-zero vectors to the range of \{0, 2\}.

When distance loss $\mathcal{L}_{self}$ converges 
near to surface boundary, the network may begin to oscillate.
At this point, SCC constraint on the projection path starts to
guide the network to change direction of gradient in next training batch, 
but not the sign,
thereby facilitating the model in learning correct SDFs.
Fig.~\ref{Figure2} (b) and (c) demonstrate that the predicted positive and negative values 
are quickly stabilized after using proposed SCC, 
and the zeros-level-set boundary is still well-preserved for surface representation.

\subsubsection{On-surface constraint with adversarial learning}
On-surface prior is another important geometric constraint to improve the 
surface reconstruction.
Ideally, the network will converge to actual surface boundary,
allowing us to accurately distinguish the inside and 
outside of a structure by SDF = 0 on the surface.
However, lacking on-surface prior can result in inaccurate or 
even wrong 
prediction in complex medical structures, such as the shapes 
with polygonal holes or noisy input.

Adversarial learning strategies have previously been investigated for point cloud reconstruction 
and 3D US sensorless reconstruction~\citep{li2019pugan,luoRecONOnlineLearning2023}.
To leverage more information, we propose encoding the 
OSC-ADL to enforce network to learn a more accurate shape. 
The ADL module, consisting of four fully connected MLP and 
Leaky ReLU activation functions, serves as an adversarial 
discriminator $\boldsymbol{D}$ for the reconstruction.  
Specifically, the actual boundary corresponds to a prior SDF of 0,
 representing on-surface constraint,
 while the SDF learned from network is considered the 'fake' value.
 The predicted signed distance values from network $\boldsymbol{f}$ are fed into 
 discriminator, and the confidence levels are generated using an attached sigmoid layer.
This enables the model to penalize points that are close to zero-level 
sets but do not belong to the surface.
 Through this strategy, the network's SDF predictions 
 are iteratively guided to accurately represent surface boundary 
 of anatomical structures, ultimately enhancing the precision of our surface 
 reconstruction methodology.

Overall, the entire network is optimized end-to-end by 
minimizing $\mathcal{L}_{\boldsymbol{G}}$ and 
 $\mathcal{L}_{\boldsymbol{D}}$  in an adversarial manner during the training stage.
 The $\mathcal{L}_{\boldsymbol{G}}$ is adversarial generator loss
 to train MLP with $\mathcal{L}_{self}$ and $\mathcal{L}_{scc}$, 
 and $\mathcal{L}_{\boldsymbol{D}}$ 
 is adversarial discriminator loss. 
 Here, we directly minimize the 
least-squared adversarial loss~\citep{Mao_2017_ICCV} 
 between predicted SDF values and zero scalar 
 field's value:

\begin{equation}
	\mathcal{L}_{\boldsymbol{G}}= \mathcal{\lambda}_{self} \mathcal{L}_{self} + \mathcal{\lambda}_{scc} \mathcal{L}_{SCC} + 
  \mathcal{\lambda}_{G} \frac{1}{2}[D(\boldsymbol{s})-1]^2 
	\end{equation}
 \begin{equation}
	\mathcal{L}_{\boldsymbol{D}}= \frac{1}{2}[D(\boldsymbol{s})^2+(D(\boldsymbol{s} ^{\prime})-1)^2]
	\end{equation}
where, ${\lambda}_{self}$, ${\lambda}_{scc}$ and ${\lambda}_{G}$ are loss weights, 
$\boldsymbol{D}(\boldsymbol{s})$ is the confidence value 
predicted by $\boldsymbol{D}$ from learned SDFs,
$\boldsymbol{G}$, $\boldsymbol{D}$ are optimized 
alternatively in the training process, 
$\boldsymbol{s}$ and $\boldsymbol{s} ^{\prime}$ denote the predicted signed distances and 
zero scalar field's value (on-surface constraint), respectively.
The size of the zero scalar field's value is defined following batch size.

\begin{figure}[t]
  \centering
  \includegraphics[width=7cm]{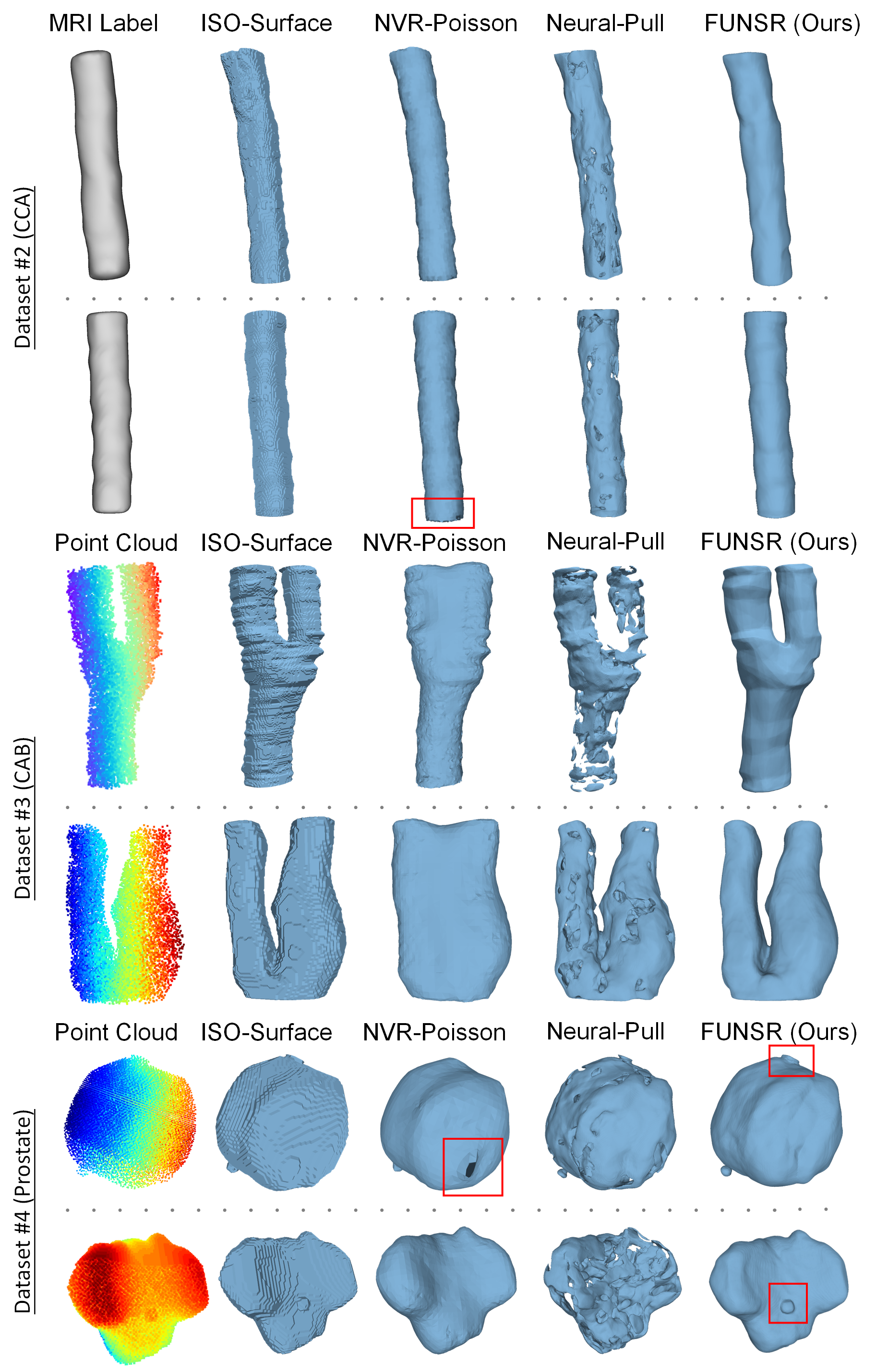}
  \caption{Six examples of reconstructed meshes from \textit{in-vivo} datasets
  using four methods
  at $1.5 \times 10^4$  iterations.
  Blue-to-red color in point cloud indicates the directional changes 
  from left to right and red boxes highlight the detailed differences between four methods.}
  \label{Figure_invivo}
\end{figure}

\section{Experiments setup}

\subsection{Datasets}
To assess the generalizability of our proposed method, 
four types of datasets are involved in this study. 
More comprehensive specifications of these datasets are listed
in~\ref{Appendix}.
\begin{itemize}
\item[1)] Phantom dataset. 
Two anthropomorphic hip joint phantoms are scanned underwater by an operator
using a portable freehand 3D US imaging system~\citep{chenCompactWirelessFreehand2020}. 
A straightforward threshold method is utilized for image segmentation. 
The corresponding computer-aided 
designed (CAD) model of each hip phantom served as a reference.

\item[2)] Common carotid artery (CCA) dataset. This dataset comprises 10 shapes of left and right CCA 
collected 
from five volunteers. 
Annotations of carotid artery are derived from an experienced rater 
through manual labeling.
Paired MRI scans for each volunteer are conducted for reference at MR Lab
at School of Biomedical Engineering at ShanghaiTech University on a 3T uMR 890 MRI system
(United Imaging Healthcare, Shanghai, China). The imaging resolution 
for CCA setting is 0.4 mm.

\item[3)] Carotid artery bifurcation (CAB) dataset. 
This dataset consists of 77 CAB shapes
collected from the clinical setting. 
Approval for this study was granted by the relevant local research ethics committees.
Reference annotations for all the images are determined through manual labeling.
We further partition the datasets into a training set (57 cases) and a test set (20 cases) 
for automatic segmentation following~\citep{zhangInvestigationInteractiveSegmentation2023}.
The automatically segmented masks from test set are utilized for 
experimental analysis of segmented noisy inputs.

\item[4)] Prostate Dataset. This dataset includes 108 patients 
from MICCAI 2023 MRI to US Registration for Prostate (µ-RegPro) Challenge~\citep{baumMRUltrasoundRegistration2023}.
The publicly available 73 transrectal ultrasound (TRUS) volumes
have been divided by organizers into 65 for training and 8 for validation in registration task.
These volumes contain well-defined 
anatomical landmarks for reference, including prostate gland, visible lesions and zonal structures etc\footnote{\url{https://muregpro.github.io/}}.
This dataset is employed to examine how our method can be directly applied to 
3D US volume with masks.

\end{itemize}

\subsection{Baseline methods}
\textbf{ISO-Surface}. This is a classic algorithm in medical imaging graphics for surface reconstruction~\citep{treeceRegularisedMarchingTetrahedra1999}.
We directly generate the surface mesh from segmented masks using ISO-Surface.

\textbf{NVR-Poisson}. Two stages are involved in this method for 
surface reconstruction of the abdominal aorta from robotic 3D US imaging~\citep{velikovaImplicitNeuralRepresentations2024}:
a) semantic neural volume reconstruction (NVR)~\citep{9958448} 
and
b) surface point cloud extraction and downsampling based on the reconstructed volume and mesh generation using 
Poisson surface reconstruction from point cloud\footnote{\url{https://www.open3d.org/docs/latest/tutorial/Advanced/surface_reconstruction.html}}.
For the implementation of stage a), 
we schedule the training of NVR for 100,000 iterations with 
a batch size of 200,000. 
Other hyperparameters are configured following~\citep{9958448},
and network architecture design is the same as~\citep{mildenhall2020nerf}.

\textbf{Neural-Pull}. Our approach is established based on the self-supervised learning strategy
that originally introduced in this method for computer vision applications~\citep{NeuralPull}.

\begin{table*}[t]
  \caption{Mean (std) results of surface reconstruction performance on different methods from four types of datasets. 
}
  \centering
  \setlength{\tabcolsep}{2mm}{}
  \begin{threeparttable}[b]
  \begin{tabular}{@{}lllllll@{}}
    \toprule
    \begin{tabular}[c]{@{}l@{}}Dataset \#1 (Phantom)\\ Methods\end{tabular}  & CC & Genus                & ASD(mm) $\downarrow$     & CD(mm) $\downarrow$      & HD(mm) $\downarrow$      & HD95(mm) $\downarrow$                \\ \midrule
    ISO-Surface                                                              & 154 (200)                                                                         & non-manifold         & 4.69 (3.41)          & 4.45 (3.22)          & 26.86 (11.48)        & 12.77(7.89)          \\
    NVR-Poisson                                                              & 14 (13)                                                                           & 99 (125)             & 4.78 (2.27)          & 4.59 (1.87)          & 26.33 (9.92)         & 13.39 (4.72)         \\
    Neural-Pull                                                              & 73 (68)                                                                           & 79 (69)              & 3.31(0.69)           & 2.78 (0.36)           & 3.43 (2.02)          & 1.75(0.18)           \\
    FUNSR(Ours)                                                              & \textbf{2 (0)}                                                                    & \textbf{1 (0)}     & \textbf{1.51(0.36)}  & \textbf{1.04 (0.01)} & \textbf{2.09 (1.59)}  & \textbf{0.77(0.38)}  \\ \midrule
    \begin{tabular}[c]{@{}l@{}}Dataset \#2 (CCA)\textsuperscript{1}\\ Methods\end{tabular}      & DSC $\uparrow$                                                                    & IoU $\uparrow$       & ASD(mm) $\downarrow$    & CD(mm) $\downarrow$     & HD(mm) $\downarrow$     & HD95(mm) $\downarrow$                 \\ \midrule
    ISO-Surface                                                              & 0.848 (0.03)                                                                       & 0.737 (0.04)          & \textbf{0.60 (0.12)}           & \textbf{0.60 (0.12)}           & \textbf{2.18(0.17)}  & 1.39 (0.29)          \\
    NVR-Poisson                                                              & 0.850 (0.02)                                                                       & 0.739 (0.04)           & \textbf{0.60 (0.11)}          & \textbf{0.60 (0.11)}          & 2.36 (0.37)          & 1.36 (0.23)          \\
    Neural-Pull                                                              & 0.662 (0.04)                                                                       & 0.496 (0.04)           & 0.64 (0.12)          & 1.03 (0.09)          & 3.99 (0.21)          & 1.48 (0.20)          \\
    FUNSR(Ours)                                                              & \textbf{0.861 (0.02)}                                                              & \textbf{0.756 (0.04)} & \textbf{0.60 (0.10)}  & \textbf{0.60 (0.10)}  & 2.31 (0.46)          & \textbf{1.22 (0.27)} \\ \midrule
    \begin{tabular}[c]{@{}l@{}}Dataset \#3 (CAB)\\ Methods\end{tabular}       & DSC $\uparrow$                                                                    & IoU $\uparrow$       & ASD(mm) $\downarrow$    & CD(mm) $\downarrow$     & HD(mm) $\downarrow$     & HD95(mm) $\downarrow$                 \\ \midrule
    ISO-Surface                                                              & 0.94 (0.02)                                                                       & 0.89 (0.04)          & \textbf{0.17 (0.03)}          & 0.16 (0.03)          & \textbf{0.75 (0.43)} & \textbf{0.28 (0.10)} \\
    NVR-Poisson                                                              & 0.82 (0.18)                                                                       & 0.72 (0.18)          & 0.62 (0.93)          & 0.64 (1.13)          & 4.04 (3.61)          & 2.05 (1.81)          \\
    Neural-Pull                                                              & 0.70 (0.19)                                                                        & 0.56 (0.17)          & 0.43 (0.90)          & 1.09 (1.90)          & 4.54 (4.31)          & 0.96 (1.92)          \\
    FUNSR(Ours)                                                              & \textbf{0.95 (0.05)}                                                              & \textbf{0.90 (0.06)} & \textbf{0.17 (0.11)} & \textbf{0.15 (0.11)} & 0.79 (0.62)          & 0.35 (0.08)          \\ \midrule
    \begin{tabular}[c]{@{}l@{}}Dataset \#4 (Prostate)\\ Methods\end{tabular}  & DSC $\uparrow$                                                                    & IoU $\uparrow$       & ASD(mm) $\downarrow$    & CD(mm) $\downarrow$     & HD(mm) $\downarrow$     & HD95(mm) $\downarrow$                 \\ \midrule
    ISO-Surface                                                              & 0.72 (0.38)                                                                       & 0.68 (0.39)          & 0.74 (0.33)          & 0.75 (0.29)          & 3.10 (4.92)          & 1.34 (1.89)          \\
    \textit{NVR-Poisson\textsuperscript{2}}                                                                 & 0.80 (0.34)                                                                       & 0.76 (0.35)          & 0.64 (0.09)          & 0.65 (0.15)          & 3.71 (2.15)          & 1.21 (0.40)          \\
    Neural-Pull                                                              & 0.79 (0.05)                                                                       & 0.66 (0.07)          & 0.63 (0.04)          & 3.12 (0.13)          & 17.17 (0.77)         & 1.19 (0.08)          \\
    FUNSR(Ours)                                                              & \textbf{0.97 (0.01)}                                                              & \textbf{0.94 (0.01)} & \textbf{0.59 (0.04)} & \textbf{0.55 (0.05)} & \textbf{1.53 (0.33)} & \textbf{1.08 (0.17)} \\ \bottomrule
    \end{tabular}
    \begin{tablenotes}\footnotesize
      \item[1] Three decimal places are used for DSC and IoU in Dataset \#2 to facilitate more nuanced comparisons.
      \item[2] Only second stage of \textit{NVR-Poisson} is involved since Dataset \#4  already contains reconstructed 3D volume.
    \end{tablenotes}
   \end{threeparttable}
    \label{table_distance_comparison}

  \end{table*}

\subsection{Evaluation metrics}

\textbf{Fidelity.} Six accuracy metrics are employed to 
comprehensively access the 
reconstruction fidelity, 
including Dice similarity coefficient (DSC), Jaccard coefficient 
(intersection over union / IoU), 
Average Surface-to-Surface Distance (ASD), 
L2-Chamfer Distance (CD),
Hausdorff Distance (HD) and 95\% Hausdorff Distance (HD95).
Here, DSC and IoU coefficients are specifically adopted to verify 
the area similarity between
 predicted closed surfaces and reference closed surfaces.
The distance metrics quantify the boundary discrepancies between 
reconstructed surfaces and their references.

\textbf{Geometry.} We also geometrically evaluate the completeness and 
smoothness of reconstructed meshes from different methods.
We choose topological measures (connected component (CC), genus~\citep{popescu-pampuWhatGenus2016,martelliIntroductionGeometricTopology2022}) to 
assess the shape completeness, and 
Gaussian curvature computation as well as kernel density estimation (KDE)
to assess the surface smoothness.

\subsection{Implementation details}

During the training phase, 
all extracted point clouds are 
randomly downsampled to \textit{N} = 20,000 using farthest point sampling (FPS)
and 25 query points are sampled for each 
point $\boldsymbol{p_i} \in \boldsymbol{P}$. 
The distance between $p_i$ and its 50-th nearest neighbor 
point $(k = 50)$ is computed to determine $\sigma$ in Section~\ref{Section_VPC_generation} for query points generation.
We use geometric network initialization strategy~\citep{atzmonSALSignAgnostic2020a} 
to initialize the weights of MLP following~\citep{NeuralPull}. 
The MLP network is initialized to approximate the signed distance function of an 
$r$-radius sphere before training. 
Here, $r$ is set to 0.5 for the bias item of MLP network.
The network is trained for $1.5 \times 10^4$ iterations 
using Adam optimizer with a learning rate of 0.001, 
a momentum of 0.9 and a batch size of 5000. 
All the segmented point clouds are normalized in a range of \{-1,1\}.
The ${\lambda}_{self}$ is set as 1, and  ${\lambda}_{scc}$, ${\lambda}_{gan}$ are 
 set as 0.005 to control the query points near to the surface, respectively.  
Our network is implemented using Pytorch and 
trained on a single NVIDIA RTX 3090 GPU with 24 GB memory.
During the reference stage, we set a threshold of Marching Cube to 0 and 
resolution to $256 \times 256 \times 256$ to generate 
mesh based on the predicted SDFs~\citep{parkDeepSDFLearningContinuous2019,NeuralPull,wiesnerGenerativeModelingLiving2024}.
This resolution choice can achieve a balance between visualization 
appearance and reference 
speed according to the findings in~\cite{wiesnerGenerativeModelingLiving2024}.

\begin{figure*}[ht]
  \centering
  \includegraphics[width=15cm]{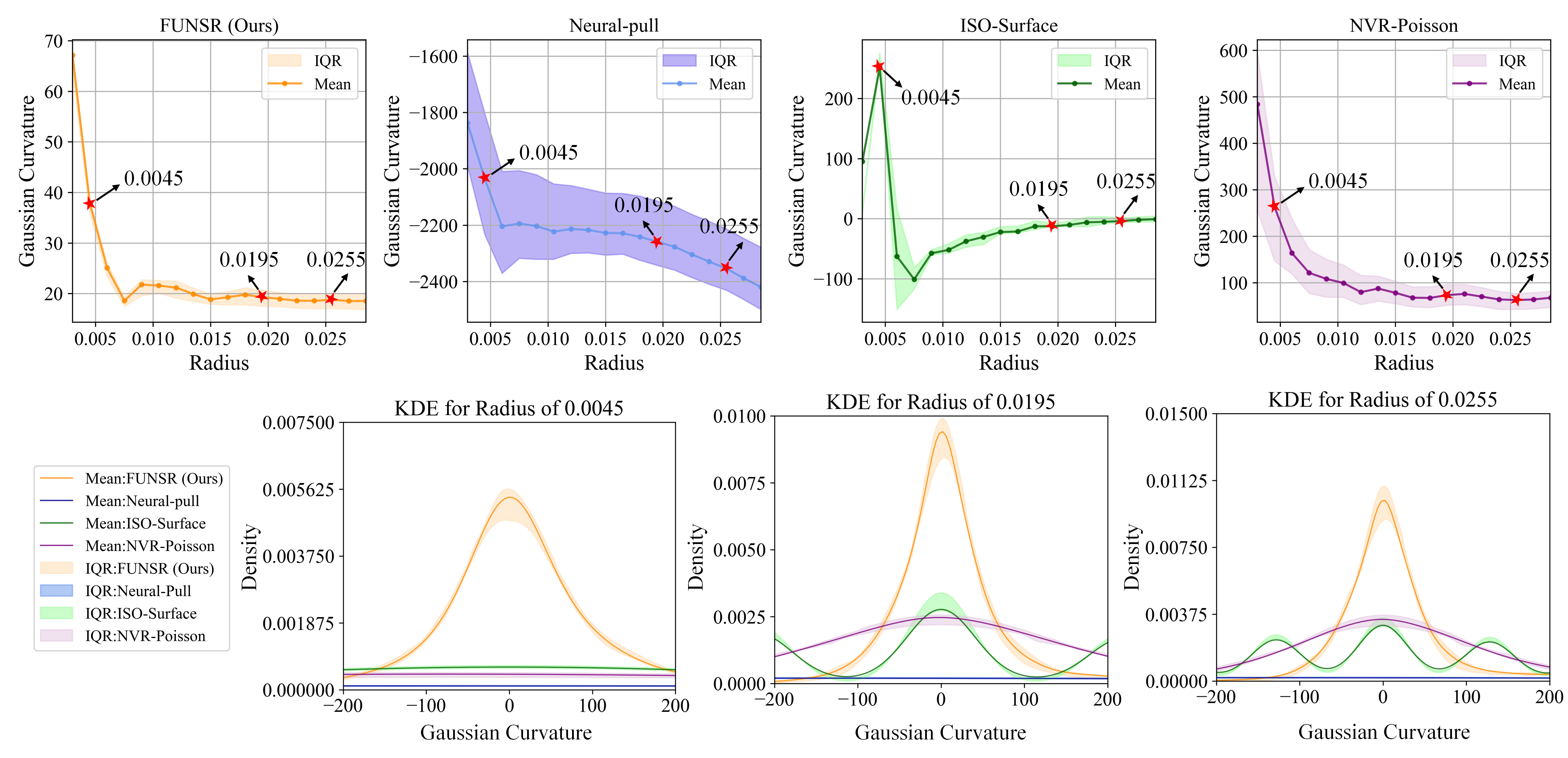}
  \caption{Quantitative evaluation of smoothness of reconstructed meshes from Dataset \#2 (CCA).
  The plots show descriptive statistics (mean and interquartile range) of four methods 
  based on Gaussian curvature (top)
  and kernel density estimation (KDE, bottom).
  For density of four methods, the higher amplitude of Gaussian curve, the smoother it is.}
  \label{Figure_invivo_Gaussian}
\end{figure*}

\section{Experimental results}

\subsection{Surface reconstruction performance}
We benchmark the structural fidelity of our results against three methods across all four datasets, 
using manual labels as inputs.
As detailed in Table~\ref{table_distance_comparison},  
our method demonstrates continuous significant superiority over three methods on Datasets \#1 and \#4,
and achieves similar performance with ISO-Surface on Dataset \#3.
Notably, we observe a substantial average decrease of 32\% ASD, 68\% in CD,
64\% in HD, and 37\% HD95 across all the datasets when compared with Neural-Pull, 
indicating the superior accuracy
of proposed method in matching reconstructed surfaces on the
respective references. 

For Dataset \#1 (phantom), we report topological measures to access the shape completeness 
rather than
similarity coefficients, as US signal can not penetrate phantom surface 
to accurately capture the structure behind surface.
The topological analysis for CAD model only has two
CC and one genus.
However, as shown at the top of Table~\ref{table_distance_comparison}, 
three baseline methods fail to maintain the accurate
shape integrity,  
particularly ISO-Surface, which results in a high number of CC 
(154 $\pm$ 200) and non-manifold shape for genus calculation.
A possible explanation for this is that Dataset \#1 contains noise and outliers introduced by threshold segmentation.
The combined NVR and Poisson surface reconstruction (NVR-Poisson) method displays some improvements over 
ISO-Surface and Neural-Pull, yet the performance is limited.
Our method (FUNSR) achieves 1.51 mm ASD
and 1.04 mm CD, approaching the imaging resolution of the data acquisition system.

\begin{figure*}[ht]
  \centering
  \includegraphics[width=15cm]{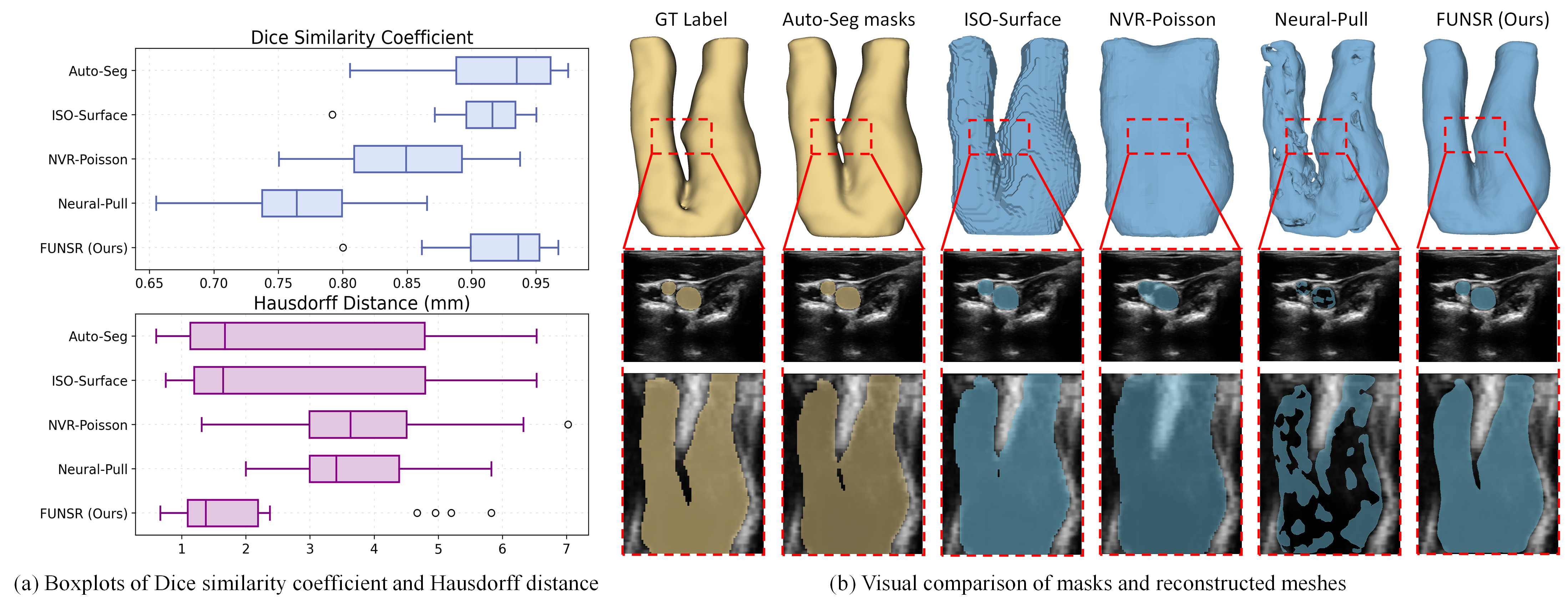}
  \caption{Quantitative and qualitative evaluation for Dataset \#3 (CAB)-test.
  (a) Boxplots of Dice similarity coefficient and Hausdorff Distance from four methods and automatic network segmentation (Auto-Seg).
  (b) Visualization of ground truth (GT) label, Auto-Seg masks and reconstructed meshes from four methods in 
  3D space, 2D cross-section plane and 2D coronal plane. 
}
  \label{Figure_seg_exp}
\end{figure*}

\begin{table*}[t]
  \centering
  \caption{Random noisy input analysis for Dataset \#4 (Prostate) at various noise level.}
  \setlength{\tabcolsep}{3mm}{}
  \begin{threeparttable}[b]
    \begin{tabular}{@{}lllllll@{}}
      \toprule
      \multicolumn{1}{c}{\multirow{2}{*}{Methods}} & \multicolumn{2}{c}{50}     & \multicolumn{2}{c}{100}     & \multicolumn{2}{c}{{200}}              \\ \cmidrule(l){2-7} 
      \multicolumn{1}{c}{}                         & HD(mm)           & CC          & HD(mm)           & CC           & HD(mm)                & CC               \\ \midrule
      ISO-Surface                                  & 36.49 (6.28) & 43.40 (4.94) & 38.34 (6.18) & 83.9 (8.48) & 39.85 (6.61)      & 165.4 (15.6)     \\
      {NVR-Poisson\textsuperscript{1,2}}                                   & 3.01 (0.88)  & \textbf{1.08 (0.39)} & 3.79 (4.07)  & 1.23 (0.44)  & \multicolumn{2}{c}{\textbackslash{}} \\
      {Neural-Pull\textsuperscript{2}}                                  & 29.23 (5.47) & 723 (2368)  & 28.14 (5.38) & 48.2 (17.4)  & 32.48 (5.68)      & 3056 (4440)      \\
      {FUNSR(Ours)\textsuperscript{2}}                                 & \textbf{2.73 (3.98)}  & \textbf{1.08 (0.40)} & \textbf{2.98 (5.19)}  & \textbf{1.08 (0.39)}  & \textbf{4.12 (7.87)}       & \textbf{1.15 (0.59)}      \\\midrule
      {FUNSR-PE\textsuperscript{3}}                               & \textbf{2.44 (3.43)}  & \textbf{1.06 (0.30)} & \textbf{2.76 (4.29)}  & \textbf{1.04 (0.20)}  & \textbf{3.28 (5.61)}       & \textbf{1.15 (0.75)}      \\ \bottomrule
      \end{tabular}
    \begin{tablenotes}\footnotesize
      \item[1] Only second stage of \textit{NVR-Poisson} is involved since Dataset \#4  already contains reconstructed 3D volume. 
      \item[2] NVR-Poisson fails to fit the convex hull at noise level 200
      and the batch size of Neural-Pull and FUNSR are set to 10000 at this noise level.
      \item[3] Reconstruction performance of FUNSR with additional Positional Encoding (PE).
    \end{tablenotes}
   \end{threeparttable}
    \label{tab_noise_study}
  \end{table*}

In Dataset \#2 (CCA), no significant differences are observed  in the results 
among ISO-Surface, 
NVR-Poisson, and our method when compared with MRI scans,
due to the simplicity of vascular structures in this dataset for reconstruction.
Similarly, ISO-Surface performs comparably to our method in Dataset \#3 (CAB). 
However, NVR-Poisson struggles to reconstruct the bifurcation structure of carotid 
artery, leading to reduced performance in this dataset.
FUNSR (Ours) demonstrates superior fidelity across the other three methods,
except for HD and HD95 metrics.
This exception can be attributed to our method's
ability to smooth out fluctuations in vascular shape caused by breathing during scans. 
Such a smoothing effect might prevent 
the reconstructed mesh from perfectly matching original input labels.

Dataset \#4 (Prostate) vividly showcases the advantages of our approach,
particularly its capability to accurately handle finer structures, 
such as tumors on the prostate gland surface (as shown in Fig.~\ref{Figure_invivo}).
In contrast, when ISO-Surface and NVR-Poisson attempt to process 
these surface details, they often create holes or lose details, 
leading to decreased DSC and IoU performance. 
A significant challenge for the second stage of NVR-Poisson is accurately 
estimating the correct point normal,
which is essential for successful Poisson surface reconstruction on these datasets.

We qualitatively present six \textit{in-vivo} examples of reconstructed meshes 
in Fig.~\ref{Figure_invivo}.
While ISO-Surface is capable of generating accurate shapes, it is 
constrained by the resolution and discontinuities.
NVR-Poisson method leverages INR and Poisson 
to produce smooth and continuous shapes. 
However, this approach may over-smooth structures in complex cases, 
for instance, 
the carotid artery bifurcations are forcibly 
closed in  Dataset\# 3 (CAB) and small raised tumors in Dataset \#4 (Prostate) 
are erased.
Additionally, Neural-Pull lacks integrity across 
all datasets due to the absence of geometric constraints.
Overall, our method demonstrates superior performance 
in both fidelity and visualization.

We further evaluate the smoothness of reconstructed meshes from different methods using Dataset \#2.
Fig.~\ref{Figure_invivo_Gaussian} shows the descriptive statistics (mean and interquartile range) of 
Gaussian curvature computation and associated kernel density estimation (KDE)
on the normalized, reconstructed mesh of each carotid artery shape.
Top plots show the variation of Gaussian curvature across different radii.
We then choose a stable radius of 0.0195 and the other two radii 0.0045 and 0.0255 
for density estimation 
at the bottom plots. 
In comparison to other three methods,
FUNSR (Ours) shows superior ability in
predicting more continuous and smooth surfaces without resolution
limitations. 
This is evidenced by the maximal probability
and minimal variation in the interquartile range.

\subsection{Various input analysis}
As previously noted, ISO-Surface exhibits the poorest performance 
on Dataset \#1 (phantom) due to unexpected noise and outliers introduced by 
simple threshold segmentation.
In this section, 
we conduct three experiments to explore the sensitivity 
of different surface reconstruction methods to different categories of input noise,
including automatic segmentation input, various random noise levels and motion perturbations.

\subsubsection{Automatic segmentation input}

Firstly, we utilize the automatic segmentation (Auto-Seg) masks of Dataset \#3 (CAB)-test
as input to assess the performance of different methods.
Fig.~\ref{Figure_seg_exp} summarizes the results both quantitatively and qualitatively.
The boxplots clearly show that
our method can partially filter out abnormal 
segmentation caused by network,
achieving the lowest HD compared to other methods.
Specifically, FUNSR exhibits a 21\% reduction in HD compared to the original Auto-Seg results.
The 3D visualization presented on the right of Fig.~\ref{Figure_seg_exp} 
aligns with boxplots,
indicating our method's efficacy in 
handling anomalies in detailed 3D structures.
Furthermore, 2D section visually illustrates the
enhanced resolution achieved by our method.

\begin{figure*}[htbp]
  \centering
  \includegraphics[width=16cm]{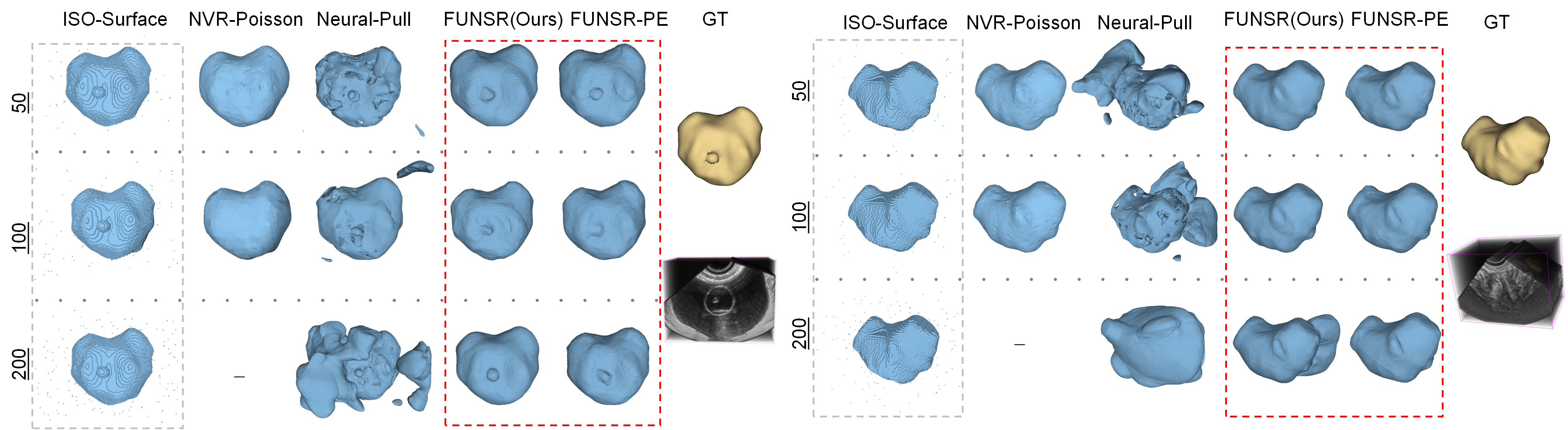}
  \caption{Visual comparisons of random noisy input analysis
  for Dataset \#4 (prostate) at noise level 50, 100 and 200. 
  All the random outliers reconstructed by ISO-Surface are bounded by gray boxes.
  Our method (FUNSR) and our method with positional encoding (FUNSR-PE) are in red boxes.
  Ground truth (GT) label and volume space are presented at the top and bottom for reference.}
  \label{Figure_denoising_exp}
\end{figure*}

\begin{figure*}[htbp]
  \centering
  \includegraphics[width=16cm]{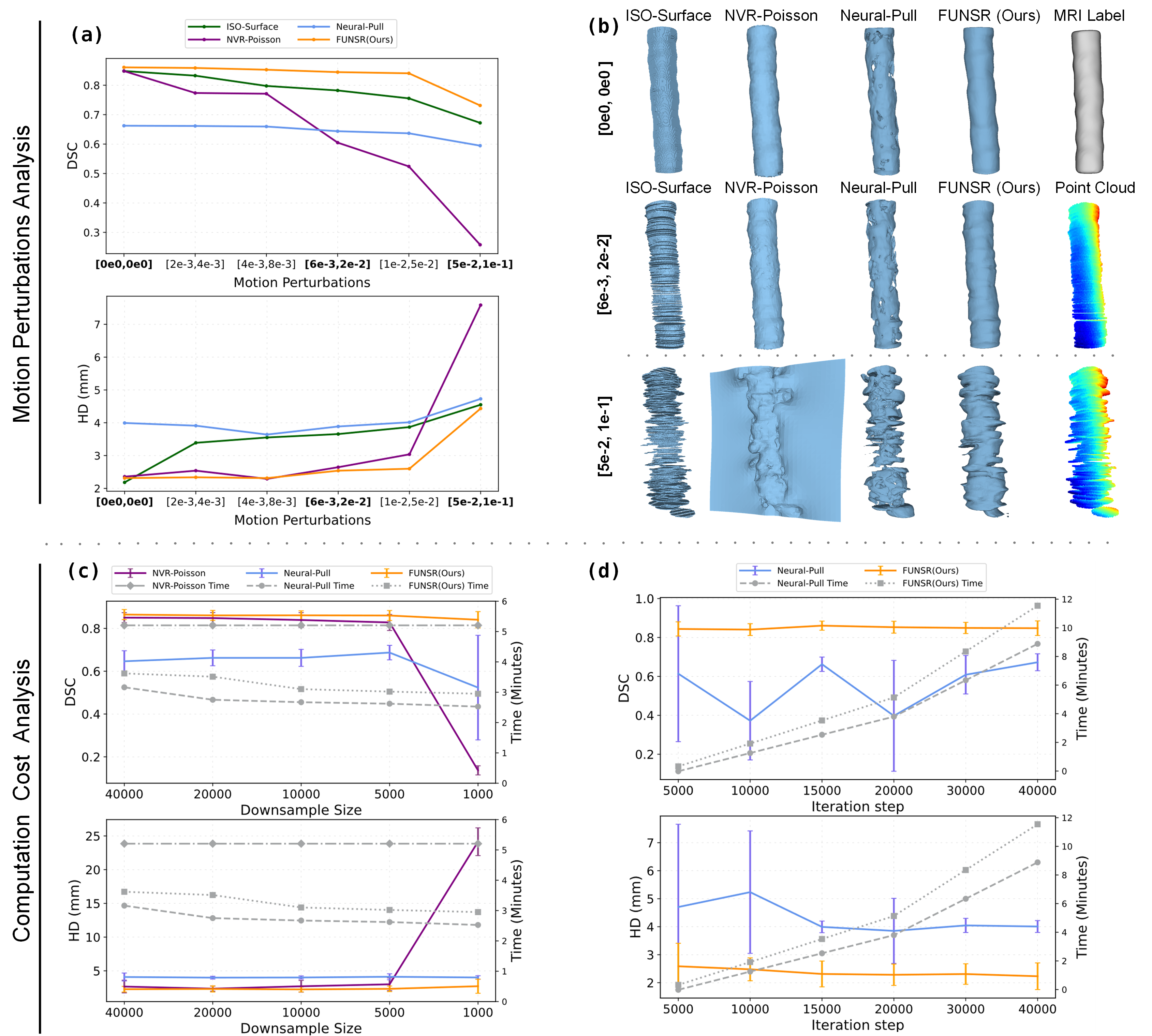}
  \caption{Quantitative and qualitative analysis of motion perturbation and computation cost for Dataset \#2 (CCA).
  (a) Plots of Dice Similarity Coefficient (DSC) and Hausdorff Distance (HD) across four methods under six perturbation levels.
  (b) Visualization of reconstructed meshes from four methods under three typical perturbation levels: 
  [$\sigma_r=0e0$, $\sigma_t=0e0$] (no motion perturbation),  [$\sigma_r=6e-3$, $\sigma_t=2e-2$], and [$\sigma_r=5e-2$, $\sigma_t=1e-1$].
  MRI label and perturbed point clouds are shown for reference.
  (c) Computation cost of various downsampling sizes for NVR-Poisson, Neural-Pull and FUNSR (Ours) at $1.5 \times 10^4$ iterations.
The time for NVR-Poisson is presented on a logarithmic scale. 
(d) Computation cost of different training iteration steps for Neural-Pull and FUNSR (Ours) with a downsampling 
size of 20,000.}
  \label{Figure_motion_time_exp}
\end{figure*}

\subsubsection{Random noise levels}

We then experiment the performance of different methods under various noise levels 
by introducing random outliers around the mask in volume space using
Dataset \#4 (prostate).
Three noise levels are introduced: 50, 100 and 200.
Two evaluation metrics that showed significant variance, 
HD and CC, are reported in Table~\ref{tab_noise_study}.
Unsurprisingly, ISO-Surface has the worst results, 
as its reconstruction directly reflects the entire structure of input.
The second stage of NVR-Poisson method 
displays competitive performance against noise interference~\citep{kazhdan2006poisson}, 
since Poisson reconstruction method typically tends to produce a closed shape.
However, this method fails to reconstruct
mesh at a high noise level, e.g., 200. 
The HD and CC of our method, with an original batch size of 5000 at this noise level, 
are 6.07 $\pm$ 10.17 mm and 1.38 $\pm$ 1.49, respectively.
While these two metrics are improved to 4.12 $\pm$ 7.87 mm and 1.15 $\pm$ 0.59 when batch size is increased to 10,000.
Moreover,
HD and CC can be further decreased across all three noise levels 
by incorporating additional positional encoding (PE)~\citep{zhengRobustPointCloud2023,tancikFourierFeaturesLet2020b}.

Fig.~\ref{Figure_denoising_exp} visually presents two examples with respect to the reconstructed meshes 
at three noise levels: 50, 100 and 200.
ISO-Surface method retains all input geometric information, 
resulting in reconstructed meshes containing all random outliers,
as indicated by gray boxes.
NVR-Poisson tends to over-smooth surface details and 
cannot fit the proper convex hull at noise level of 200 for these two cases.
Both our method and our method with PE demonstrate superiority in addressing random noisy inputs, 
as highlighted by red boxes.
On the right of Fig.~\ref{Figure_denoising_exp},
we show a failure case of our method at noise level of 200,
where the network is unable to correctly restore clear shape under such high-level noise. 
However, positional encoding can aid our network in 
fully reconstructing the actual prostate structure.

\subsubsection{Motion perturbation}

We further analyze the influence of random perturbation on transducer's motion during 
freehand scan using Dataset \#2 (CCA).
We separately perturb the rotation component ($\mathbf{R} \in SO(3) $) and 
translation component ($\mathbf{t}\in\mathbb{R}^3$) of each pose with additive noise 
$ \boldsymbol n_{r} $ and  $ \boldsymbol n_{t} $
following~\cite{espositoTotalVariationRegularization2019} and~\cite{chenLocaltoGlobalRegistrationBundleAdjusting2023}.
Here, $\boldsymbol n_{r}$ and $\boldsymbol n_{t}$ are elements of  $\mathfrak{s e}(3)$ Lie algebra, 
with $\boldsymbol n_r \sim \mathcal{N}(\mathbf{0}, \sigma_{r} \mathbf{I})$ 
and $\boldsymbol n_t \sim \mathcal{N}(\mathbf{0}, \sigma_{t} \mathbf{I})$.
The motion perturbation is represented in 6-DOF (degree-of-freedom) and formulated
as $\mathbf{T} =[\mathbf{R} \, | \, \mathbf{t}] \in SE(3) $. 
Here, $SO(3)$ and $SE(3)$ stand for the Special Orthogonal (SO) group  
and Special Euclidean (SE) group in three dimensions, respectively.

Fig.~\ref{Figure_motion_time_exp}(a) demonstrates the performance degradation of four methods under various motion 
perturbations ranging from  
0 (no perturbation) to [$\sigma_r = 5e-2$, $\sigma_t = 1e-1$], 
which corresponds to a maximum absolute deviation of 9.86 degrees in rotation
and 0.32 pixels in translation.
It is evident that ISO-Surface 
is highly susceptible to interference from motion, 
as shown by 
HD plot and visual comparison in 
Fig.~\ref{Figure_motion_time_exp}(b).
The fragments and holes in the reconstructed mesh 
from NVR-Poisson method
lead to a serious decrease in DSC starting at $[\sigma_r = 6e-3, \sigma_t = 2e-2$],
and yielding the worst DSC and HD at level $[\sigma_r = 5e-2, \sigma_t = 1e-1$].
To plot this level, we exclude the cases where the convex hull cannot be fit for mesh generation.
Consistent with previous experiments, 
Neural-Pull losses 
complete shape contour across all input categories.
Our proposed FUNSR presents competitive reconstruction quality 
compared to other three methods
and shows ability to handle the largest level of perturbation at 
$[\sigma_r = 1e-2, \sigma_t = 5e-2$].
The value of HD eventually increases to a similar discrepancy 
as ISO-Surface and 
Neural-Pull at $[\sigma_r = 5e-2, \sigma_t = 1e-1$].
Fig.~\ref{Figure_motion_time_exp}(b) qualitatively 
shows three typical perturbation levels 
as well as corresponding reconstructed meshes from four methods.
The comparisons highlight our advantage of noise robustness.

\subsection{Computation cost analysis}
Differing from ISO-Surface, 
which directly reconstructs the mesh from segmented mask, 
other three methods involve downsampling operation for the network input. 
In this section, therefore, we explore the impact of downsampling size on reconstruction performance and 
computation cost
using Dataset \#2 (CCA).
The results are plotted in Fig.~\ref{Figure_motion_time_exp} (c).
The time required for NVR-Poisson method shows no variation 
across different downsampling sizes, 
with an average duration of 182.3 minutes. 
This suggests that the computational cost for this method is 
primarily driven by training of semantic-NVR.
Although the proposed geometric constraints introduced a 
slight computational burden than Neural-Pull, 
the total processing time is still within the accepted range, about 3.5 minutes.
Besides, Fig.~\ref{Figure_motion_time_exp} (c) indicates that 
20,000 points can achieve a balance between performance and time for our method
under the current setting of query point generation.

Additionally, we investigated the performance of Neural-Pull and our 
method at different numbers of training iterations, as well as the computational time.
Fig.~\ref{Figure_motion_time_exp} (d) shows the unstable convergence of Neural-Pull at different iteration steps, 
in contrast to consistent performance exhibited by our approach.
Meanwhile,  the plots reveal that training for $1.5 \times 10^4$ steps of our method achieves a balance between
the reconstructed quality and time cost.
Moreover, a considerable performance can be rapidly produced at 5000 iterations, which takes less than 1 minute.
The above results significantly demonstrate the 
ability of our method to realize high reconstruction quality and 
fast convergence speed.

\subsection{Ablation Study}
In this subsection, 
to investigate the capability of each module in proposed method, 
we perform a comprehensive ablation study on Dataset \#1 (phantoms), which has CAD model as
ground truth.
The experiments include exploring the effectiveness of two geometric constraints
and performance of point cloud modalities.
We train the network for $2.0 \times 10^4$ 
iterations and compare the results of each module at $1.5 \times 10^4$ and $2.0 \times 10^4$ iterations.
Neural-Pull is chosen as the baseline method.

\begin{table*}[ht]
  \caption{Ablation studies on the effectiveness of geometric constraints using Dataset \#1 (phantoms)  
  at $1.5 \times 10^4$ and $2.0 \times 10^4$ iterations.}
  \centering
  \setlength{\tabcolsep}{1.1mm}{}
  \begin{tabular}{@{}llclc@{}}
    \toprule
    \multirow{2}{*}{Module} & \multicolumn{2}{c}{$1.5 \times 10^4$}                                          & \multicolumn{2}{c}{$2.0 \times 10^4$}                                         \\ \cmidrule(l){2-5} 
                             & CD (mm)                                & \multicolumn{1}{l}{HD (mm)}           & CD (mm)                               & \multicolumn{1}{l}{HD (mm)}           \\ \midrule
    Baseline                 & 2.78(0.36)                             & 3.43(2.02)                            & 3.29(0.38)                            & 4.21(1.44)                            \\
    +SCC                      & 1.26(0.09)                             & 2.31(1.57)                            & 1.29(0.16)                            & 2.34(0.95)                            \\
    +OSC-ADL                  & 1.67(0.11)                             & 2.67(1.39)                            & 1.75(0.54)                            & 2.65(0.84)                            \\
    FUNSR(Ours)              & \textbf{1.04 (0.01)} & \textbf{2.09(1.59)} & \textbf{1.23(0.15)} & \textbf{1.97(1.01)} \\ \bottomrule
    \end{tabular}
    \label{tab_ablation_study1}
  \end{table*}

  \begin{figure*}[t]
    \centering
    \includegraphics[width=10.5cm]{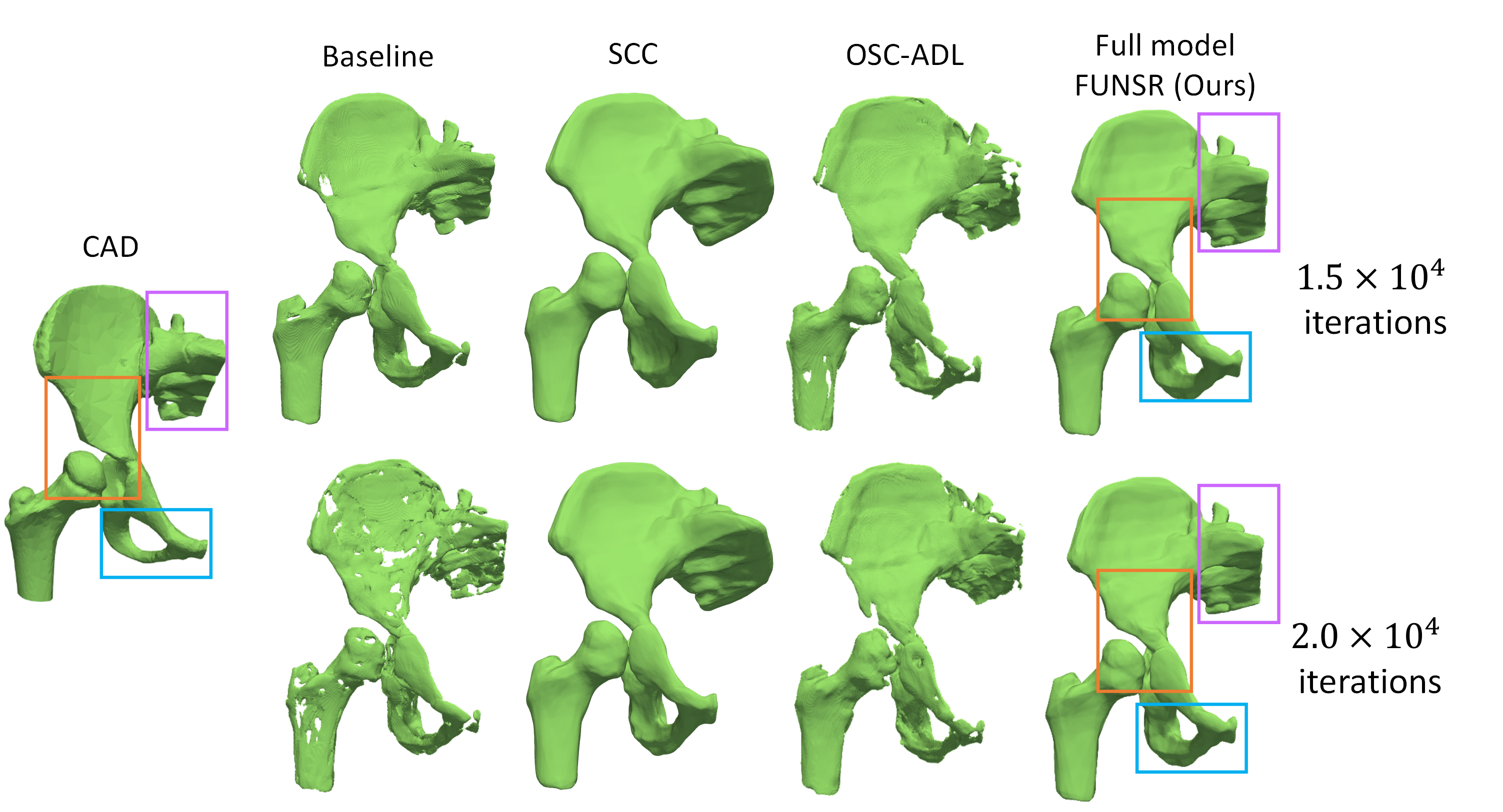}
    \caption{Visualization of the ablation study on
    the H1 phantom in Dataset \#1 for validation of geometric constraints. 
    The top and bottom plots are reconstructed surfaces of baseline (Neural-pull),
    only SCC, only OSC-ADL
    and full model (FUNSR) at $1.5 \times 10^4$ and $2.0 \times 10^4$ iterations, respectively.
    The CAD model is shown on the left.
    Three colored boxes highlight the prominent differences.}
    \label{Figure_ablation_study}
  \end{figure*}
  
  \begin{figure*}[h]
    \centering
    \includegraphics[width=12cm]{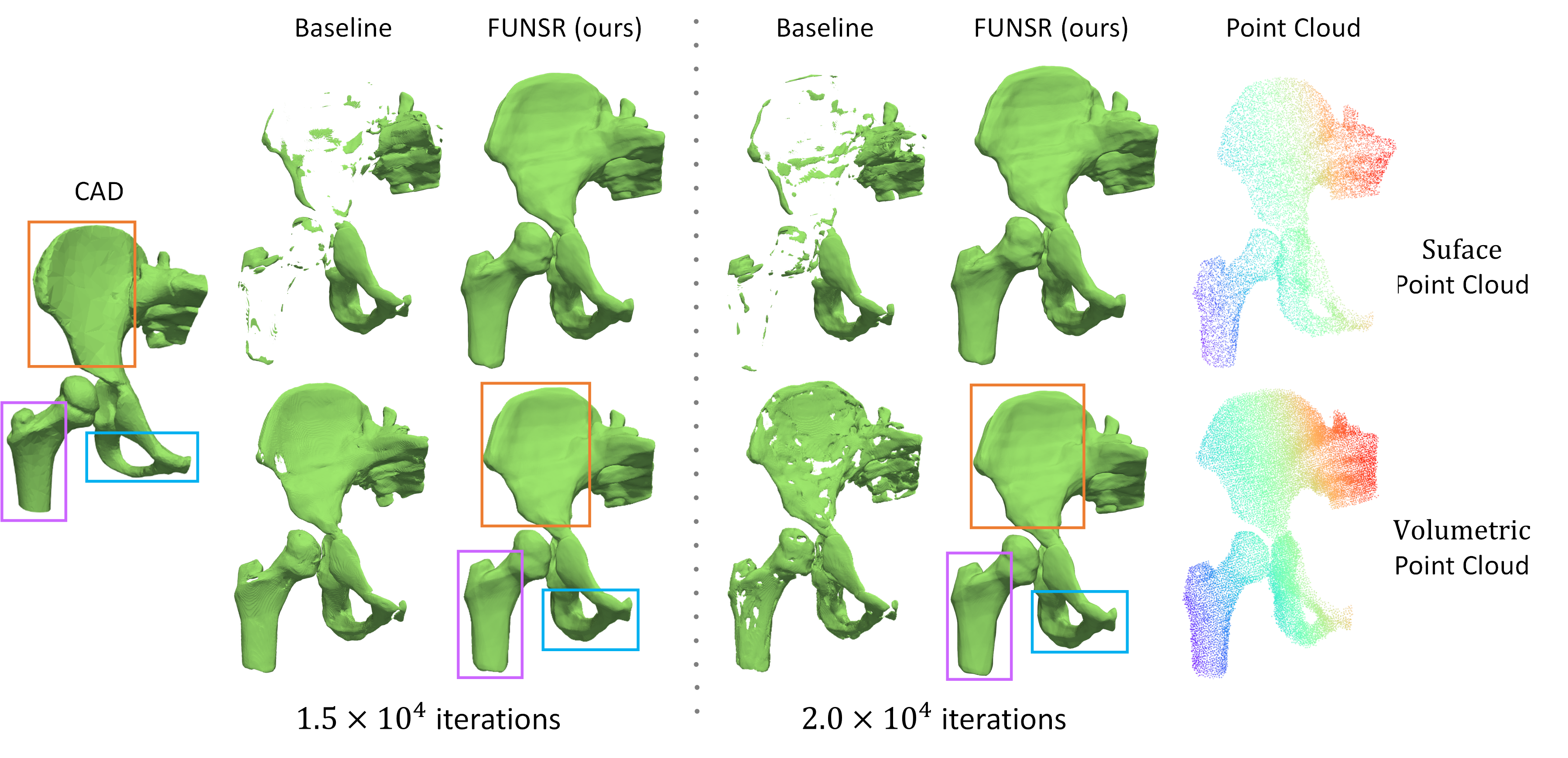}
    \caption{Visualization of the ablation study on
    the H1 phantom in Dataset \#1 for comparison of different inputs. 
    The results at $1.5 \times 10^4$ and $2.0 \times 10^4$ iterations are presented
    on the left block and right block.
     The top and bottom plots are reconstructed surface of baseline (Neural-pull)
    and our method using surface point cloud and volumetric point cloud, respectively.
    The CAD model is shown on the left. 
    Blue-to-red color in point
    cloud indicates the directional changes from left to right.
    Three colored boxes highlight the prominent differences.}
    \label{Figure_ablation_study2}
  \end{figure*}

\subsubsection{Validating the effectiveness of geometric constraints}
The effectiveness of our method is 
attributed to the incorporation of two geometric constraints, 
SCC and OSC-ADL. 
We have individually evaluated their 
efficiency on two hip phantom models  
through both qualitative 
and quantitative analysis.
The ablation study is carried out by
separately adding SCC and OSC-ADL to the baseline.

Table~\ref{tab_ablation_study1} reports individual contribution of SCC and OSC-ADL 
to the enhancement of surface reconstruction performance.
Compared with baseline at $1.5 \times 10^4$ iterations,
the SCC results improve by 54\% for CD and 33\% for HD.
OSC-ADL also demonstrates a decrease of
 40\% for CD and 22\% for HD.
Furthermore, the most significant improvements are observed when both 
constraints are jointly optimized in proposed method,
revealing a lower CD (63\%) and HD (39\%) on two phantoms.
Our method can also improve the accuracy
by decreasing 63\% CD, 53\% HD
at $2.0 \times 10^4$.

\begin{table*}[h]
  \caption{Ablation studies on the performance of two input point cloud modalities, surface point cloud (S) versus 
  volumetric point cloud (V), using Dataset \#1 (Phantom)
  at $1.5 \times 10^4$ and $2.0 \times 10^4$ iterations. }
  \renewcommand\arraystretch{1}  
  \centering
  \setlength{\tabcolsep}{1mm}{}
  \begin{tabular}{@{}llccc@{}}
    \toprule
    \multirow{2}{*}{Methods} & \multicolumn{2}{c}{$1.5 \times 10^4$}                                         & \multicolumn{2}{c}{$2.0 \times 10^4$}                                         \\ \cmidrule(l){2-5} 
                             & CD (mm)                               & HD (mm)                               & CD (mm)                               & HD (mm)                               \\ \midrule
    Baseline-S               & 12.35(2.39)                            & 24.20(1.52)                            & 13.26(1.06)                            & 23.78(2.34)                            \\
    Baseline-V                   & 2.78(0.36)                            & 3.43(2.02)                            & 3.29(0.38)                            & 4.21(1.44)                            \\
    Proposed-S                   & 1.12(0.06)                            & 2.17(1.58)                            & 1.43(0.41)                            & 2.07(0.92)                             \\
    Proposed-V                   & \textbf{1.04(0.01)} & \textbf{2.09(1.59)}  & \textbf{1.23(0.15)} & \textbf{1.97(1.01)} \\ \bottomrule
    \end{tabular}
  \label{tab_ablation_study2}
  \end{table*}

Fig.~\ref{Figure_ablation_study} shows an example of
ablation study conducted on H1 phantom model.
Baseline method presents the worst structures, 
showing the broken surface and imperfect shape due to unstable training, especially at 
$2.0 \times 10^4$ iterations. 
SCC constraint illustrates the ability to reconstruct overall 
shape at both iterations compared to the baseline.
However, only using SCC fails to handle certain tough 
detailed structures at $1.5 \times 10^4$ iterations, 
as indicated by the purple and blue boxes.
The edge of the sacrum (purple) appears to be inflated and 
obturator foramen at the bottom (blue) is filled.
In contrast, OSC-ADL module can recover the detailed structures, 
but lose the completeness of general morphology due to the absence of SCC.
By combining both constraints, 
our proposed method is able to 
effectively preserve all the structures.

 \subsubsection{Comparing the appearance of the VPC with SPC}
In this experiment, we investigate our network performance using surface point cloud (SPC) or volumetric point cloud (VPC) 
as input. 
The SPC data, which serves as the original input for baseline,
is extracted only from segmented boundary.
While the VPC data is generated 
from all the voxels within segmentation masks. 
A visual contrastive analysis of two input modalities is shown on the right of
Fig.~\ref{Figure_ablation_study2}.
The VPC
exhibits a dense aggregation of 3D points within the mask area, which provides
enriched information for underlying shape representation.
Therefore, only using SPC as input 
will lead the network to produce a coarser local surface or even lose integrity 
as opposed to 
the results using VPC.
Main differences are most apparent on the three highlight boxes.
Our method shows the advantage of representing
continuity and completeness, particularly in achieving a smoother surface 
when using VPC, as indicated by the orange and blue boxes in Fig.~\ref{Figure_ablation_study2}.

We also present a quantitative assessment using CD and HD 
in Table~\ref{tab_ablation_study2}.
The numerical results align with visual inspection in Fig.~\ref{Figure_ablation_study2},
with baseline showing that SPC input yields the highest CD and HD 
due to the vanishing boundaries in reconstructed mesh.
In contrast, the performance of our method utilizing 
SPC or VPC is both significantly improved in comparison to baseline
and achieves the lowest discrepancy when using VPC.

\section{Discussion and Conclusion}
We propose FUNSR, an end-to-end self-supervised surface reconstruction model 
designed to tackle the challenges in 3D freehand US,
including the limited resolution and noise-segmented boundary for 
representation of anatomical structure.
Although the proposed method proves to be valuable for the current situation, 
it's important to acknowledge limitations and discuss future work to address them.
First, even though the reconstruction time for individual subject is already 3.5 minutes 
by using a single NVIDIA RTX 3090 GPU, 
the computational efficiency can still be improved for possible real-time reconstruction~\citep{kerbl3DGaussianSplatting2023}.
For simplicity, we directly consider the entire segmentation mask 
as volumetric input in this study.
Future studies could benefit from investigating resampling strategies that concentrate 
on sampling more points around the boundary and fewer points at the center of the mask to 
accelerate reconstruction process further.
Second, the input of VPC will result in a pseudo-SDF
field inside the shape.
While the primary goal of this paper is to accurately reconstruct surface 
structure and simultaneously generate an external signed distance field for possible navigation applications, 
a more precise internal distance field and inner structure could be achieved with assistance of multi-view US scans or 
unsigned distance functions~\citep{chenNeuralImplicitRepresentation2023}.

In summary, we have developed a novel DL-based approach
that incorporates geometric constraints
 to learn 
 SDFs of neural implicit surface reconstruction for
freehand 3D US volume. 
The proposed constraints allow us to effectively minimize the sign variations and 
reconstruction errors during the training.  
The results demonstrate our method is highly effective at 
reconstructing the 3D surface of medical anatomical 
structures.
Additionally, we show that our approach can be directly applied to 3D US volumes
for surface reconstruction.
The promising results have the potential to advance applications of 3D
US in medical augmented reality, surgical navigation, and computer-assisted interventions.

\appendix
\renewcommand\thetable{\Alph{section}\arabic{table}}    
\section{}
\setcounter{table}{0}
\label{Appendix}

\begin{table*}[t]
  \centering
  \setlength{\tabcolsep}{3mm}{} 
  \renewcommand\arraystretch{1.2} 
  \caption{Appendix A: Overview of datasets, data acquisition systems, grid size and data segmentation methods in the experiments.}
  \label{table_appendix}
  \footnotesize
  \centering
  \begin{threeparttable}[b]
  \begin{tabular}{@{}llccccc@{}}
    \toprule
    Data & Sweep & \multicolumn{1}{l}{\begin{tabular}[c]{@{}l@{}}Ultrasound\\ Transducer\end{tabular}} & \multicolumn{1}{l}{Frequecy} & \multicolumn{1}{l}{\begin{tabular}[c]{@{}l@{}}Tracking \\ Device\end{tabular}} & \multicolumn{1}{l}{\begin{tabular}[c]{@{}l@{}}Grid\\ Size\end{tabular}} & \multicolumn{1}{l}{Segment}                                        \\ \midrule
    \#1  & 2     & \begin{tabular}[c]{@{}c@{}}Clarius, C3HD \\Canada\end{tabular}                    & 5 MHz                        & \begin{tabular}[c]{@{}c@{}}Polhemus \\ G4, U.S.A.\end{tabular}                 & \begin{tabular}[c]{@{}c@{}}0.5 \\ (mm)\end{tabular}                                                                     & Threshold                                                               \\ \specialrule{0em}{1pt}{2pt}
    \#2  & 10    & \begin{tabular}[c]{@{}c@{}}Clarius, L7HD \\ Canada\end{tabular}                    & 10 MHz                       & \begin{tabular}[c]{@{}c@{}}Polhemus \\ G4, U.S.A.\end{tabular}                & \begin{tabular}[c]{@{}c@{}}0.2 \\ (mm)\end{tabular}                                                                & \begin{tabular}[c]{@{}c@{}}Manual\\ Labeling \\\end{tabular}              \\ \specialrule{0em}{1pt}{2pt}

    \#3  & 77    & \begin{tabular}[c]{@{}c@{}}Clarius, L7HD\\ Canada\end{tabular}                    & 10 MHz                       & \begin{tabular}[c]{@{}c@{}}Polhemus \\ G4, U.S.A.\end{tabular}                & \begin{tabular}[c]{@{}c@{}}0.2 \\ (mm)\end{tabular}                                                                 & \begin{tabular}[c]{@{}c@{}}Network \& \\ Manual \\ Labeling\\\end{tabular} \\
    \#4  & 73    & \begin{tabular}[c]{@{}c@{}}Clinic TRUS \\ Machine\end{tabular}                   & 9 MHz                        & -                                                                              & \begin{tabular}[c]{@{}c@{}}0.8 \\ (mm)\end{tabular}                                                               & \begin{tabular}[c]{@{}c@{}} Organizer\end{tabular}        \\ \bottomrule
    \end{tabular}
    \begin{tablenotes}\footnotesize
      \item[*] Grid size is set following ~\citep{chenCompactWirelessFreehand2020,liAutomaticDiagnosisCarotid2023, zhangInvestigationInteractiveSegmentation2023,baumMRUltrasoundRegistration2023}
      for Dataset  \#1, Dataset  \#2, Dataset  \#3 and  Dataset  \#4, respectively.
    \end{tablenotes}
   \end{threeparttable}
  \end{table*}

Table~\ref{table_appendix} summarizes the datasets and detailed technical parameters
of used data acquisition systems.
The spacing size of voxel grid 
is set according to the pixel size of US transducer 
based on previous works~\citep{chenCompactWirelessFreehand2020,liAutomaticDiagnosisCarotid2023, zhangInvestigationInteractiveSegmentation2023,baumMRUltrasoundRegistration2023}.

\section*{Acknowledgement}
This research was supported 
by the Natural Science Foundation of 
China (NSFC) under Grant No.12074258 and a grant from Natural Science Foundation of 
China (NSFC) under Grant No.62071299, and
a grant from Alberta Innovates - Accelerating Innovations into CarE (AICE) program
under Grant No. RES0056222.

\section*{CRediT authorship contribution statement}
\textbf{Hongbo Chen:} Writing - Original Draft, 
Conceptualization (Equal), Methodology (Equal), Visualization,
Software. 
\textbf{Logiraj Kumaralingam:}Writing - Review \&
Editing, Conceptualization (Equal), Methodology (Equal),
Validation (Equal). 
\textbf{Shuhang Zhang:} Visualization, Validation. 
\textbf{Sheng Song:} Data Curation, Software. 
\textbf{Fayi Zhang:} Resources, Software. 
\textbf{Haibin Zhang:} Resources, Data Curation. 
\textbf{Thanh-Tu Pham:} Data Curation. 
\textbf{Kumaradevan Punithakumar:} Writing - Review \&
Editing. 
\textbf{Edmond H. M. Lou:} Investigation, Data Curation.
\textbf{Yuyao Zhang:} Data Curation, Funding Acquisition.
\textbf{Lawrence H. Le:} Writing
- Review \& Editing, Funding Acquisition. 
\textbf{Rui Zheng:} Writing - Review \& Editing, Funding Acquisition,
Supervision.

\bibliographystyle{model2-names.bst}\biboptions{authoryear}
\bibliography{Surface-Reconstruction}

\end{document}